\documentclass[final,numberedheadings]{aipproc}%
\usepackage{amsmath}
\usepackage{amsfonts}
\usepackage{amssymb}
\usepackage{graphicx}%

 \layoutstyle{6x9}

\begin{document}

  \noindent {USC-06/HEP-B3
  \footnote{{\small {Lectures delivered at
  SUSY06, the 14th International Conference on Supersymmetry and the Unification
  of Fundamental Interactions, Irvine, CA, June 2006, and at} { the
  26$^{th}$ International Colloquium on Group Theoretical
  Methods in Physics, New York, NY, June 2006. Transparencies available at
  http://physics.usc.edu/\symbol{126}bars/papers/2TSMlecture.pdf.}
  \small  }
  }
  \hfill\hfill hep-th/0610187}

 \title{The Standard Model as a 2T-physics Theory}


\author{Itzhak Bars
}{
  address={Department of Physics and
  Astronomy\\ University of Southern California, Los Angeles, CA
  90089-0484, USA}
  }

\begin{abstract}
New developments in 2T-physics, that connect 2T-physics field theory
directly to the real world, are reported in this talk. An action is
proposed in field theory in 4+2 dimensions which correctly
reproduces the Standard Model (SM) in 3+1 dimensions (and no junk).
Everything that is known to work in the SM still works in the
emergent 3+1 theory, but some of the problems of the SM get
resolved. The resolution is due to new restrictions on interactions
inherited from 4+2 dimensions that lead to some interesting physics
and new points of view not discussed before in 3+1 dimensions. In
particular the strong CP violation problem is resolved without an
axion, and the electro-weak symmetry breakdown that generates masses
requires the participation of the dilaton, thus relating the
electro-weak phase transition to other phase transitions (such as
evolution of the universe, vacuum selection in string theory, etc.)
that also require the participation of the dilaton. The underlying
principle of 2T-physics is the local symmetry Sp(2,R) under which
position and momentum become indistinguishable at any instant. This
principle inevitably leads to deep consequences, one of which is the
two-time structure of spacetime in which ordinary 1-time spacetime
is embedded. The proposed action for the Standard Model in 4+2
dimensions follows from new gauge symmetries in field theory related
to the fundamental principles of Sp(2,R). These gauge symmetries
thin out the degrees of freedom from 4+2 to 3+1 dimensions without
any Kaluza-Klein modes. The extra 1+1 dimensions are compensated by
the Sp(2,R) gauge symmetry that removes ghosts and cure other
problems such as causality. The SM emerges from 4+2 dimensions as
one of the gauge choices in coming down from 4+2 to 3+1 dimensions.
As is usual in 2T-physics, there are many ways of embedding 3+1 in
4+2 as gauge choices, and this should lead to holographic images
that appear as different 1T-dynamics, but are dual field theories of
the SM. This is likely to lead to new methods for investigating QCD
and other field theories. The dualities among the 3+1 dimensional
images, and the hidden symmetries of 4+2 dimensions realized by each
image, is part of the evidence for the underlying 4+2 dimensions.
\end{abstract}

 \maketitle


\section{Sp$\left( 2,R\right) $ gauge symmetry and 2T-physics}

The essential ingredient in 2T-physics is the basic gauge symmetry
Sp$(2,R)$ acting on phase space $X^{M},P_{M}$ \cite{2treviews}.
Under this gauge symmetry, momentum and position are locally
indistinguishable, so the symmetry leads to some deep consequences
\cite{2treviews} - \cite{Dirac}. One of the strikingly surprising
aspects of 2T-physics is that a given $d+2$ dimensional 2T theory
descends, through Sp$\left( 2,R\right) $ gauge fixing, down to a
family of holographic 1T images in (d-1)+1 dimensions, all of which
are gauge equivalent to the parent 2T theory and to each other.
However, from the point of view of 1T-physics each image appears as
a different dynamical system with a different Hamiltonian. Some of
the phenomena that emerge in 2T-physics include certain types of
dualities, holography and emergent spacetimes, which have been
studied primarily at the level of classical and quantum particle
dynamics (Fig.1).
 \begin{center}
 \fbox{\includegraphics[ height=4.4944in, width=5.9819in]%
 {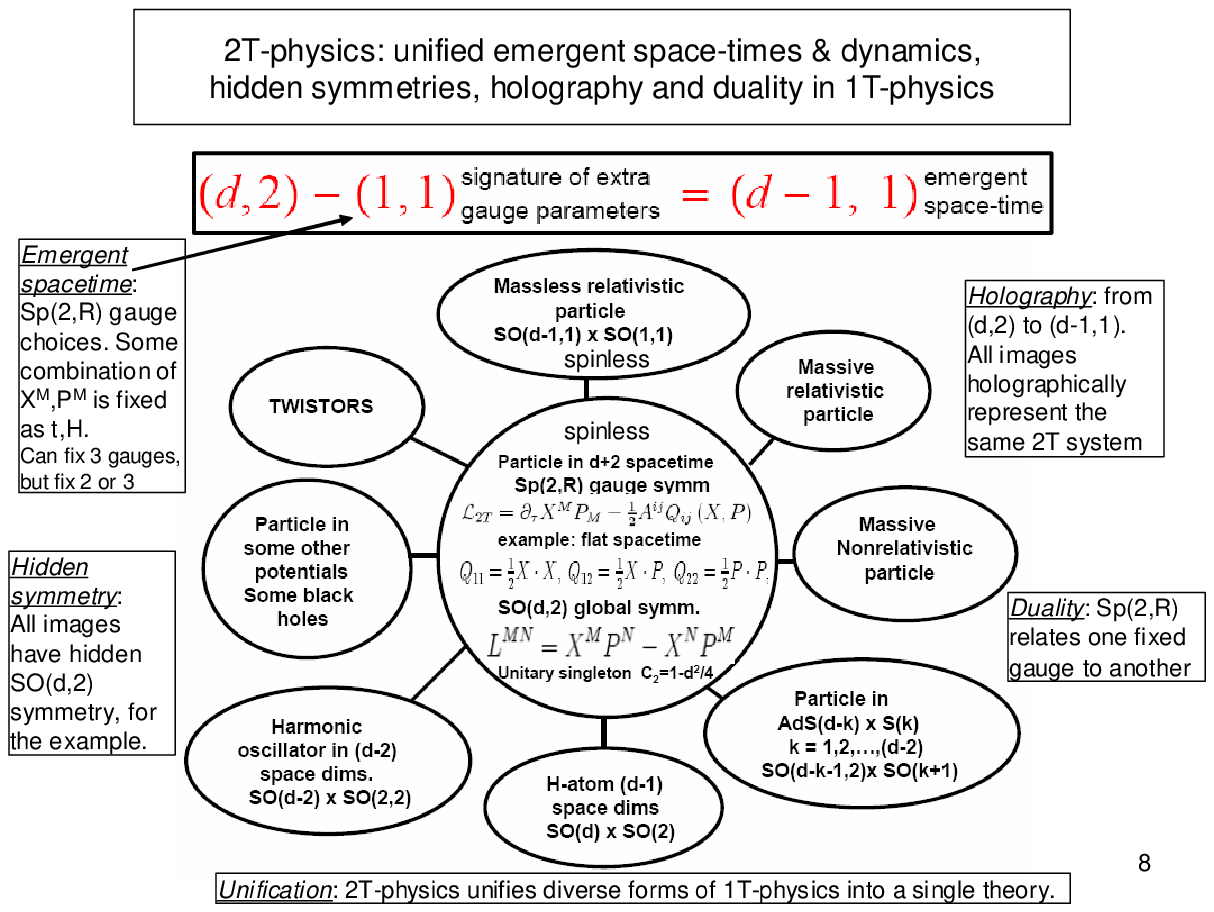}}\\
 Fig.1 - Some 1T-physics systems that emerge from the solutions of Q$_{ij}=0.$%
 \end{center}
 Hence, 2T-physics is a
unification approach for 1T-physics systems through higher
dimensions. It is distinctly different than Kaluza-Klein theory
because there are no Kaluza-Klein towers of states, but instead
there is a family of 1T systems with duality type relationships
among them.

In the field theoretic 4+2 SM the type of phenomena summarized in
Fig.1 are expected. For the time being only one of the field
theoretic images, namely the one labeled as \textquotedblleft
massless relativistic particle\textquotedblright\ in Fig.1, has been
studied. It is this 3+1 holographic image of the 4+2 SM that
coincides with the well known 3+1 SM. This emergent SM has some
attractive features: it solves the long standing CP problem, and
leads to novel ideas on the origins of mass generation, as mentioned
in the abstract.

I now describe the main features of the 2T-physics field theory and
its consequences in $4+2$ dimensions. The generic field theory with
fields of
spins $0,1/2,$ and $1$ includes a set of SO$\left( 4,2\right) $ vectors $%
A_{M}^{a}\left( X\right) $ labeled with $M=$ SO$\left( 4,2\right) $
vector, and $a=$ the adjoint representation of some Yang-Mills gauge
group $G;$
scalars $H_{i}\left( X\right) ,$ labeled by an internal symmetry index $%
i=1,2,\cdots $ (a collection of irreducible representations of $G$);
left or
right handed spinors $\Psi _{L\alpha }^{I}\left( X\right) ,\Psi _{R\dot{%
\alpha}}^{\tilde{I}}\left( X\right) $ in the $4,4^{\ast }$
representations of SU$\left( 2,2\right) =$SO$\left( 4,2\right) ,$
labeled with $\alpha
=1,2,3,4,$ and $\dot{\alpha}=1,2,3,4,$ and internal symmetry indices $%
I=1,2,\cdots $ and $\tilde{I}=1,2,\cdots $ (again, a collection of
irreducible representations of $G$). The generic Lagrangian has the
form of a Yang-Mills theory in $4+2$ dimensions ($G$-covariant
derivatives), except for space-time features shown explicitly in the
Lagrangian below, needed to impose the underlying Sp$\left(
2,R\right) $ gauge symmetry and the related 2T-physics gauge
symmetries.

There is no space here to give the details of the 2T-physics gauge
symmetries in field theory \cite{2tbrst2006,2tstandardM}, but I note
the basic important fact that the equations of motion that follow
from the Lagrangian below impose the Sp$\left( 2,R\right) $ gauge
singlet conditions $X^2=X\cdot P=P^2=0$ indicated in Fig.1, but now
including interactions \cite{2tstandardM}
\begin{align*}
L& =\delta \left( X^{2}\right) \left\{ -D_{M}H^{i\dagger
}D^{M}H_{i}\right\}
+2~\delta ^{\prime }\left( X^{2}\right) ~H^{i\dagger }H_{i} \\
& +\delta \left( X^{2}\right) \left\{
\begin{array}{c}
\frac{i}{2}\left( \overline{\Psi _{L}}^{I}X\bar{D}\Psi
_{IL}+\overline{\Psi
_{L}}^{I}\overleftarrow{D}\bar{X}\Psi _{IL}\right)  \\
-\frac{i}{2}\left( \overline{\Psi _{R}}^{\tilde{I}}\bar{X}D\Psi _{\tilde{I}%
R}+\overline{\psi _{R}}^{\tilde{I}}\overleftarrow{\bar{D}}X\Psi _{\tilde{I}%
R}\right)
\end{array}%
\right\}  \\
& +\delta \left( X^{2}\right) \left\{ g_{I}^{i\tilde{I}}\overline{\Psi _{L}}%
^{I}X\Psi _{\tilde{I}R}H_{i}+\left( g_{I}^{i\tilde{I}}\right) ^{\ast
}H^{\ast i}\overline{\Psi _{R}}^{\tilde{I}}\bar{X}\Psi _{IL}\right\}  \\
& +\delta \left( X^{2}\right) \left\{ -\frac{1}{4}F_{MN}^{a}F_{a}^{MN}-V%
\left( H,H^{\ast },\Phi \right) \right\}  \\
& -\frac{1}{2}\delta \left( X^{2}\right) ~\partial _{M}\Phi \partial
^{M}\Phi +\delta ^{\prime }\left( X^{2}\right) ~\Phi ^{2}
\end{align*}%
The distinctive space-time features in 4+2 dimensions include the
delta function $\delta \left( X^{2}\right) $ and its derivative
$\delta ^{\prime }\left( X^{2}\right) $ that impose
$X^{2}=X^{M}X_{M}=0,$ the kinetic terms of fermions that include the
factors $X\bar{D},\bar{X}D,$ and Yukawa couplings that include the
factors $X$ or $\bar{X},$ where $X\equiv \Gamma ^{M}X_{M},$
$\bar{D}=\bar{\Gamma}^{M}D_{M}$ etc., with $4\times 4$ gamma
matrices $\Gamma ^{M},\bar{\Gamma}^{M}$ in the $4$,$4^{\ast }$
spinor bases of SU$\left( 2,2\right) $=SO$\left( 4,2\right) .$ This
Lagrangian is not invariant under translation of $X^{M},$ but is
invariant under the spacetime rotations SO$\left( 4,2\right) .$

This Lagrangian has precisely the right space-time, and gauge
invariance, properties for the $4+2$ field theory to yield the usual
$3+1$ field theory via gauge fixing, with the usual kinetic terms
and Yukawa couplings in the emergent $3+1$ dimensional Minkowski
space $x^{\mu }$. The emergent $3+1$ theory contains just the right
fields as functions of $x^{\mu }$: all extra degrees of freedom
disappear without leaving behind any Kaluza-Klein type modes or
extra components of the vector and spinor fields in the extra 1+1
dimensions. Furthermore, the emergent $3+1$ field theory is
invariant under translations and Lorentz transformations SO$\left(
3,1\right) $:  these Poincar\'{e} symmetries are included in
SO$\left( 4,2\right) $ that takes the non-linear form of conformal
transformations in the emergent $3+1$ dimensional space-time $x^{\mu
}$, as indicated in Fig.1.

As in the last line of the Lagrangian, one may also include an additional SO$%
\left( 4,2\right) $ scalar, the dilaton $\Phi \left( X\right) ,$
classified as a singlet under the group $G.$ The dilaton is not
optional if the action is written in $d+2$ dimensions (see
\cite{2tstandardM}), as it appears in
overall factors $\Phi ^{\frac{2\left( d-4\right) }{d-2}},\Phi ^{-\frac{d-4}{%
d-2}}$ multiplying the Yang-Mills kinetic term and Yukawa terms
respectively, in order to achieve the 2T-gauge symmetry of the action. In $%
4+2$ dimensions ($d=4$) these factors reduce to $1,$ but the dilaton
can still couple to the scalars $H$ in the potential $V\left(
H,H^{\ast },\Phi \right) .$

While almost all of the usual terms of $3+1$ dimensional Yang-Mills
theory coupled to matter appear in the $3+1$ theory that emerges
from the $4+2$ field theory above, there are two notable exceptions
that play an important and interesting physical role when we apply
the $4+2$ approach to construct the Standard Model. Namely,

\begin{itemize}
\item The 2T-gauge symmetry requires the potential $V\left( H,H^{\ast },\Phi
\right) $ to be purely quartic, i.e. no mass terms are permitted.
Then the emergent 3+1 theory cannot have mass terms for the scalars,
and is automatically invariant under scale transformations.

\item There is no way to generate a term in the emergent $3+1$ theory that
is analogous to the P and CP-violating $F_{\mu \nu }F_{\lambda
\sigma }\varepsilon ^{\mu \nu \lambda \sigma }$ term that is
possible in $3+1$
dimensions. Its absence\footnote{%
In searching for possible 4+2 sources that may generate the CP
violating term in 3+1, one must include the requirement that the
$4+2$ theory should not include terms that descend to
non-normalizable interactions in $3+1.$ Actually there appears as if
there would be a topological term of the form $\int
d^{6}X~\varepsilon ^{M_{1}M_{2}M_{3}M_{4}M_{5}M_{6}}Tr\left(
F_{M_{1}M_{2}}F_{M_{3}M_{4}}F_{M_{5}M_{6}}\right) $ whose density is
a total divergence for any Yang-Mills gauge group $G,$ and therefore
should not affect the equations of motion, or renormalizability.
However, as discussed in \cite{2tstandardM}, this term can be gauge
fixed to zero by using the 2T gauge symmetry. In this sense, the 2T
gauge symmetry plays a similar role to the Peccei-Quinn symmetry in
eliminating the topological term. But one must realize that the 2T
gauge symmetry is introduced for other more fundamental reasons and
also it is not a global symmetry. Hence, unlike the Peccei-Quinn
symmetry it does not lead to an axion.} is due to the fact that the
Levi-Civita symbol in $4+2$ dimensions has 6 indices rather than
$4,$ and also due to the combination of 2T gauge symmetry as well as
Yang-Mills gauge symmetry. The absence of this CP-violating term is
of crucial importance in the axionless resolution of the strong CP
violation problem of QCD suggested in \cite{2tstandardM}.
\end{itemize}

The 2T-physics field theory above is applied to construct the
Standard Model
in 4+2 dimensions by choosing the gauge group $G=$SU$\left( 3\right) \times $%
SU$\left( 2\right) \times $U$\left( 1\right) $ and including the
usual matter representations for the Higgs, quarks and leptons
(including right handed neutrinos in singlets of $G$), but now as
fields in $4+2$ dimensions. As explained in \cite{2tstandardM} this
theory descends to the usual Standard Model in $3+1$ dimensions.

The emergent $3+1$ theory leads to phenomenological consequences of
considerable significance. In particular, the higher structure in
$4+2$ dimensions prevents the problematic $F\ast F$ term in QCD.
This resolves the strong CP problem without a need for the
Peccei-Quinn symmetry or the corresponding elusive axion.

Mass generation with the Higgs mechanism is less straightforward
since the tachyonic mass term is not allowed. However by taking the
potential of the form $V\left( \Phi ,H\right) =\frac{\lambda
}{4}\left( H^{\dagger }H-\alpha ^{2}\Phi ^{2}\right) ^{2}$ we obtain
the breaking of the electroweak symmetry by the Higgs doublet
$\langle H\rangle $ driven by the vacuum expectation value of the
dilaton $\langle \Phi \rangle $, thus relating the the two phase
transitions to each other. In this way the 4+2 formulation of the
Standard Model provides an appealing deeper physical basis for mass.
In addition, there are some brand new mechanisms of mass generation
related to the higher dimensions that deserve further study (e.g.
massive particle gauge in Fig.1).

The dilaton driven electroweak phase transition makes a lot more
sense conceptually than the usual approach in which the electroweak
phase transition is an isolated phenomenon. This is because the
Higgs vacuum expectation value fills all space everywhere in the
universe. This is a hard concept to swallow without relating it to
the evolution of the universe, which then requires the participation
of gravity. By having $\langle H\rangle $ being driven by the
dilaton, not only the relation to the behavior of the gravity
multiplet is established, but also a relation is established to
other phase transitions, such as the vacuum selection process in
string theory (which depends on the dilaton), and perhaps even to
inflation that is driven by a scalar field which could be the
dilaton.

Although unclear at the present for lack of a full understanding of
the quantum theory, we may have also found the seeds for a
resolution of the hierarchy problem in 2T-physics. Namely, the
absence of quadratic terms is a consequence of symmetry at the
classical level, and if the symmetry is not anomalous it would lead
to the absence of the quadratic divergence at the quantum level,
thus resolving the hierarchy problem.\footnote{I thank C. Johnson
for drawing my attention to this point.}

The 4+2 formulation requires the kinds of concepts above that are
not required and therefore not contemplated in the literature of the
usual Standard Model or its extensions in $3+1$ dimensions.

The presence of the dilaton, as well as the right handed neutrinos,
are features with further phenomenological consequences. Both of
these are very weakly coupled to standard matter. Hence they are
possible candidates for Dark Matter. In particular, the dilaton
communicates with all other matter only through the Higgs. If a
Higgs is found at the LHC, it could provide a gateway to measure
phenomenological effects of the required dilatonic degree of freedom
in the 4+2 formulation of the Standard Model.

In summary, I emphasize the following physics points that the new
formulation of the Standard Model implies

\begin{itemize}
\item 2T-physics works and is a physical theory! Local Sp$(2,R)$ (i.e. $X,P$
indistinguishable) is a fundamental principle that agrees with
everything we know about Nature as embodied by the Standard Model.

\item The Standard Model in 4+2 dimensions provides new guidance to
fundamental physics: it resolves the strong CP violation problem of
QCD, and leads to the appealing concept of dilaton driven
electroweak spontaneous symmetry breakdown.

\item A weakly coupled dilaton may have phenomenological signals at the LHC.
It could also be a candidate for Dark Matter. Since this is a
required particle in the 4+2 formulation, and since this formulation
solves the fundamental CP problem, the dilaton concept and its
possible observation at the LHC must be taken seriously.

\item The 2T-physics concepts are easily applied to physics Beyond the Standard Model,
including GUTS, SUSY, and gravity. All of these can be elevated to
2T-physics in d+2 dimensions. These will be reported in the near
future.

\item Strings and branes have only been partially formulated in 2T-physics.
In particular, tensionless strings and branes, and the twistor
superstring, have already been obtained.

\item Similarly, it is expected that 2T-physics is a good guide for
constructing M-theory in 11+2 dimensions with an OSp$(1|64)$ global
SUSY.

\item There seems to be practical advantages of formulating 1T physics from
the vantage point of d+2 dimensions. This is seen by examining the
type of phenomena summarized in Fig.1. Namely, 2T-physics provides
new insights: emergent spacetimes and dynamics, unification,
holography, duality, hidden symmetries, that could be used to
analyze theories such as QCD and others in a non-perturbative
fashion. I hope that this aspect of 2T-physics will become an
effective tool in the future.
\end{itemize}

\section*{Acknowledgments}

I gratefully acknowledge discussions with S-H. Chen, Y.C. Kuo, B.
Orcal, and G. Quelin. This research was supported by the US
Department of Energy, under Grant No. DE-FG03-84ER40168

\end{document}